# An End-to-End Mispronunciation Detection System for L2 English Speech Leveraging Novel Anti-Phone Modeling

*Bi-Cheng Yan[1,2], Meng-Che Wu[2], Hsiao-Tsung Hung[2], Berlin Chen[1]*

[1] National Taiwan Normal University, Taiwan
[2] ASUS AICS, Taiwan

{bicheng_yan, meng_wu, alexht_hung}@asus.com, berlin@ntnu.edu.tw

## Abstract

Mispronunciation detection and diagnosis (MDD) is a core component of computer-assisted pronunciation training (CAPT). Most of the existing MDD approaches focus on dealing with categorical errors (viz. one canonical phone is substituted by another one, aside from those mispronunciations caused by deletions or insertions). However, accurate detection and diagnosis of non-categorial or distortion errors (viz. approximating L2 phones with L1 (first-language) phones, or erroneous pronunciations in between) still seems out of reach. In view of this, we propose to conduct MDD with a novel end-to-end automatic speech recognition (E2E-based ASR) approach. In particular, we expand the original L2 phone set with their corresponding anti-phone set, making the E2E-based MDD approach have a better capability to take in both categorical and non-categorial mispronunciations, aiming to provide better mispronunciation detection and diagnosis feedback. Furthermore, a novel transfer-learning paradigm is devised to obtain the initial model estimate of the E2E-based MDD system without resource to any phonological rules. Extensive sets of experimental results on the L2-ARCTIC dataset show that our best system can outperform the existing E2E baseline system and pronunciation scoring based method (GOP) in terms of the F1-score, by 11.05% and 27.71%, respectively.

**Index Terms**: computer-assisted pronunciation training (CAPT), mispronunciation detection and diagnosis (MDD), end-to-end ASR, anti-phone model

## 1. Introduction

Computer-assisted pronunciation training (CAPT) systems provide opportunities of self-directed language learning for second-language (L2) learners. It can supplement the teachers' instructions, offer individualized feedback and also mitigate the problem of teacher shortage. The mispronunciation detection and diagnosis (MDD) module play an integral role in CAPT systems, since this module facilitates to pinpoint mispronunciation segments and provide phone-level diagnosis feedback.

The MDD methods developed so far can be roughly grouped into two categories. The first is pronunciation scoring based methods, which compute phone-level pronunciation scores based on confidence measures derived from ASR, e.g., phone durations, phone posterior probability scores and segment duration scores [1], [2], [3]. Goodness of pronunciation (GOP), based on the log-likelihood ratio test, and its variants are the most representative methods of this category. However, these methods typically can only provide the functionality of mispronunciation detection, but lack the ability of providing appropriate mispronunciation diagnosis. The second category of methods aims to assess the details of mispronunciations, providing diagnosis feedback about specific errors such as phone substitutions, deletions and insertions [4], [5], [6]. A well-known method of this category is the extend recognition network (ERN) method, which extends the decoding network of ASR with phonological rules and thus can readily provide diagnosis feedback based on comparison between an ASR output and the corresponding text prompt. Nevertheless, on one hand, it is difficult to enumerate and include sufficient phonological rules into the decoding network for all L1-L2 language pairs. On the other hand, inclusion of too many phonological rules would degrade ASR accuracy, thereby leading to poor MDD performance. More recently, the end-to-end (E2E) based ASR paradigm instantiated with connectionist temporal classification (CTC) [7] has also been introduced to MDD to with promising results, in comparison to the GOP-based method that builds on the hybrid deep neural network-hidden Markov model (DNN-HMM) based acoustic model [2]. Among others, there also has been some follow-up work of using disparate E2E-based methods to address the MDD problem [8], [9].

However most of the aforementioned methods have focused exclusively on detecting categorical pronunciation errors (e.g., phoneme substitutions, insertions or deletions), whereas paying less attention to detecting mispronunciations that belong to non-categorial or distortion errors [10], [11]. As an illustration, Figure 1 shows the MDD results of a mispronounced utterance of an L2 English speaker, where the yellow blocks correspond to mispronunciation segments. The canonical phone-level pronunciation for word "The" is [dh iy] but an L2 speaker uttered [d ah] instead, where [dh]→[d] and [iy]→[ah] are categorical errors (viz., substitutions). In addition, the canonical pronunciation of the consonant in word "He" should be [hh], but it is instead pronounced as [hh*], which in fact is a non-categorial pronunciation error.

In view of this, we propose to approach MDD with a specially-tailored E2E-based ASR model structure, where the involved E2E-based model is embodied with a hybrid CTC-Attention model [12]. By combining the strengths of both CTC and the attention-based model, it is anticipated that the resulting composite model can utilize CTC to assist the attention-based model to compensate for the misalignment problem and improve the speed of the decoding process, though the attention-based model provides flexible soft-alignment between the output label sequence and the input acoustic feature vector sequence without any Markov assumptions as CTC does. Furthermore, we expand the original L2 phone set with their corresponding anti-phone set, making the proposed E2E-based MDD approach have the ability to take in both categorical and non-categorial mispronunciations, in order to provide better mispronunciation detection and diagnosis feedback [13]. Furthermore, a novel transfer learning paradigm is devised to obtain the initial model estimate of the E2E-based ASR system without resource to any phonological rules. The rest of the paper is organized as follows. We first elucidate the model architectures of the proposed E2E-based MDD methods in Section 2, followed by the experimental setup and results in Section 3. Finally, we conclude the paper and suggest avenues for future work in Section 4.

| Transcription | He | Obeyed | | The | Pressure | | | | Of | Her | Hand | |
|---|---|---|---|---|---|---|---|---|---|---|---|---|
| Canonical | hh iy | ow b ey | d | dh iy | p r eh sh er | | | | ah v | hh er | hh ae n | d |
| Annotation | #hh iy | ow b ey | t | d ah | p r ae sh er | | | | ah v | sil sil | hh #aa n | d |
| CTC | hh iy | ow b sil | d | d ah | p sil ae sh er | | | | ah v | sil sil | hh #aa n | d |
| Attention | hh iy | ow b ey | d | dh ah | p r ae sh ah | | | | ah v | hh er | hh er n | d |
| CTC-Attention | hh iy | ow b ey | d | ah #r | p r ae sh er | | | | ow v | sil er | hh ae n | d |

Figure 1: *Analysis of the results generated by different baseline E2E-based MDD methods.*

## 2. End-to-End MDD Model

In this section, we first describe the hybrid CTC-Attention model that we capitalize on for E2E-based MDD. After that, we explain the notion of anti-phone modeling that will be realized for E2E-based MDD. Then, we shed light on the trick for estimating the initial anti-phone probabilities and the full training procedure for the E2E-based model.

### 2.1. CTC/Attention-based Modeling Architecture

We adopt a hybrid CTC-Attention model (or CTC-ATT for short) architecture, originally designed for E2E-based ASR, to tackle the MDD problem [12]. In this architecture, the attention-based model takes the primary role in determining output symbols, through the use of effective attention mechanisms to perform flexible alignment between an input acoustic vector sequence and the associated output symbol sequence. On the other hand, CTC, normally sharing the encoder with the attention-based model, makes use the Markov assumptions to alleviate irregular alignment between input acoustic vector sequence and the output symbol sequence. CTC plays an auxiliary role here to assist the attention-based model for more accurate MDD performance. The CTC-Attention model will predict an $L$-length phone sequence $Y = y_1..y_l..y_L$ (e.g., $y_l$ belongs to the standard IPA symbol set) given a $T$-length input acoustic feature vector sequence $O = \mathbf{o}_1..\mathbf{o}_t..\mathbf{o}_T$. In the context of MDD, the output phone sequence Y can be viewed as the diagnosis result, in relation a text prompt that corresponds to O. For the attention-based model, the probability distribution $P_{att}(Y|O)$ is computed by multiplying the sequence of conditional probabilities of label $y_l$ given the past history $y_{1:l-1}$:

$$P_{att}(Y|O) = \prod_{l=1}^{L} P_{att}(y_n|y_{1:l-1}, O). \quad (1)$$

Subsequently, $P_{att}(y_l|O, y_{1:l-1})$ is obtained with the joint encoder and decoder networks. The encoder network can be a bidirectional long short-term memory (BLSTM) which extracts a high-level hidden acoustic vector sequence $H^E = (\mathbf{h}_1^E, ..., \mathbf{h}_S^E)$ from the input acoustic feature vector sequence O:

$$H^E = \text{BLSTM}(O), \quad (2)$$

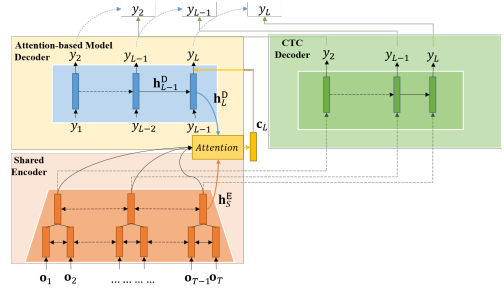

Figure 2: *A schematic depiction of the* hybrid *CTC-Attention model architecture for MDD.*

where $S$ is the length of the hidden vector sequence and usually $S < T$ due to the subsampling operation. The decoder network is a disparate unidirectional long short-term memory network (ULSTM), which predicts the incoming phone label $y_l$ conditioning on the previous output $y_{l-1}$, the current decoder state $\mathbf{h}_l^D$ and the context vector $\mathbf{c}_l$:

$$y_l = \text{Softmax}\left(\text{LinB}\left(\text{LSTM}(y_{l-1}, \mathbf{c}_l, \mathbf{h}_l^D)\right)\right), \quad (3)$$

$$\mathbf{h}_l^D = \text{ULSTM}(cat(y_{l-1}, \mathbf{c}_l), \mathbf{h}_{l-1}^D), \quad (4)$$

where $\text{LinB}(\cdot)$ is a linear transformation. The input of the ULSTM in (4) consists of the previous decoder's state $\mathbf{h}_{l-1}^D$ and the concatenation of $\mathbf{c}_l$ and $y_{l-1}$. The context vector $\mathbf{c}_l$ can be calculated using an attention mechanism which communicates information between the encoder's holistic representation $H^E$ and the current decoder's states $\mathbf{h}_l^D$. The attention mechanism is summarized as follows:

$$\mathbf{c}_l = \sum_{s=1}^{S} \mathbf{a}_{l,s} \mathbf{h}_s^E, \quad (5)$$

$$\mathbf{a}_{l,s} = \text{Align}(\mathbf{h}_l^D, \mathbf{h}_s^E),$$
$$= \frac{\exp\left(\text{Score}(\mathbf{h}_l^D, \mathbf{h}_s^E)\right)}{\sum_{s=1}^{S} \exp\left(\text{Score}(\mathbf{h}_l^D, \mathbf{h}_s^E)\right)}. \quad (6)$$

The soft-alignment (association) between a hidden acoustic vector state and a decoder state is quantified with a normalized score function $\text{Score}(\cdot,\cdot)$; here we adopt the location-based scoring function [14].

On a separate front, CTC first generates a frame-wise symbol sequence $\mathbf{z} = z_1..z_l..z_T$. The probability of an output symbol sequence Y compute by:

$$P_{ctc}(Y|O) = \sum_{\mathbf{z}} \prod_{S} P(z_s|z_{s-1}, Y)P(z_s|O)P(Y), \quad (7)$$

where $P(z_s|z_{s-1}, Y)$ represents the state transition probability, which satisfies the monotonic alignment constraint posed by CTC. In the context of MDD, the inclusion of $P(z_s|z_{s-1}, Y)$ can bring benefit to the MDD task, since the model will learn transitions between mispronunciations and correct pronunciations from the training corpus. $P(z_s|O)$ is the frame-level label probability and computed by

$$P(z_s|O) = \text{Softmax}\left(\text{LinB}(\mathbf{h}_s^E)\right). \quad (8)$$

In the training phase, the loss of CTC and the loss of the attention-based model are combined with an interpolation weight $\lambda \in [0, 1]$, so as to encourage monotonic alignments.

$$\mathcal{L}_{hyb} = \lambda \mathcal{L}_{ctc} + (1 - \lambda)\mathcal{L}_{att}. \quad (9)$$

The hybrid CTC-Attention model architecture is also adopted in the test phase. The additional incorporation of the CTC

objective is expected to provide fast and accurate inference during the training and test phases, thanks to its monotonic-alignment property. Figure 2 shows a schematic depiction of the hybrid CTC-Attention model architecture for MDD.

### 2.2. Creation of the Anti-phone Set

In order to model non-categorical errors, we introduce the notion of anti-phones to the CTC-Attention model architecture, which is designed to accommodate non-categorical mispronunciations of each L2 phone. To create an anti-phone set, each phone symbol in the L2 canonical phone set $\mathcal{U}^{can}$ is appended with a token # at its beginning to designate its anti-phone to be added into the anti-phone set $\mathcal{U}^{anti}$. As such, the resulting augmented phone symbol set $\mathcal{U}$ for the E2E-based MDD model will be the union of the canonical phone set and the anti-phone set:

$$\mathcal{U} = \mathcal{U}^{can} \cup \mathcal{U}^{anti}. \quad (10)$$

Taking advantage of this augmented phone symbol set, it is anticipated that the associated E2E-based MDD model can separate mispronunciations into categorical errors and non-categorical errors. In this way, for a mispronunciation that is in between a L2 (target) canonical phone and some L1 (mother-tongue) phone pronunciations, or is a distortion of the canonical phone pronunciation, it would be possible to detect and classify this mispronunciation with the associated anti-phone label of the canonical phone.

### 2.3. Data Augmentation with Label Shuffling

In this subsection, we describe a novel data-augmentation process for E2E-based anti-phone modeling, which creates additional speech training data with a label-shuffling scheme. Specifically, for every utterance in the original speech training dataset, the label of a phone $\varphi$ at each position of its reference transcript is either kept unchanged or randomly substituted with an arbitrary anti-phone label (excluding the anti-phone label that corresponds to $\varphi$) with a predefined probability. As such, we can duplicate the original speech training data, having the new copy be equipped with the label-shuffled transcripts that contain anti-phone labels. Note here that "the original speech training dataset" mentioned above refers to the part of non-native English utterances in the training dataset of the L2-ARCTIC corpus [15] (L2:CP; *cf.* Section 3.1) that were correctly pronounced without any pronunciation errors.

### 2.4. Training of the E2E-based MDD Model

The training process of the proposed E2E-based model for MDD can be broken down into three stages. At the first stage, an accent-free E2E model is trained on a publicly-available English speech dataset that contain utterances of native speakers (which was compiled from the TIMIT corpus and a small portion of the Librispeech corpus [16]; *cf.* Section 3.1). The second stage is to train an accent-contained E2E model. To this end, we first adopt the notion of transfer learning to initialize the encoder network of the accent-contained E2E model with the corresponding parameters of the accent-free E2E model trained at the first stage [17]. Then, the decoder network of the accent-contained E2E model is trained on the augmented dataset (containing only of non-native English training utterances without mispronunciations) described in Section 2.3, while the encoder network of the accent-contained E2E model is finetuned with this augmented dataset. Finally, at the third stage, the whole accent-contained E2E model is finetuned with the rest L2 English training utterances that contain mispronunciations (L2:MP; *cf.* Section 3.1). Note also that for a mispronounced phone segment of a given training utterance, its phone label in the transcript of the utterance is replaced with its corresponding anti-phone label. We argue that the aforementioned training procedure could enable the resulting accent-contained E2E model not only to identify and diagnose categorical mispronunciation errors accurately, but also to detect non-categorical mispronunciations to some extent.

## 3. Experiments

### 3.1. Speech Corpora and Model Architecture

We used the L2-ARCTIC[1] corpus for our experiments [15], which is a publicly-available non-native English speech corpus intended for research in mispronunciation detection, accent conversion, and others. This corpus contains correctly pronounced utterances (denoted by CP) and mispronounced utterances (denoted by MP) of 24 non-native speakers, whose L1 languages are Hindi, Korean, Mandarin, Spanish, Arabic and Vietnamese. We divided each of these two parts of utterances into training, development and test subsets, respectively. As mentioned in Section 2.4, a suitable amount of native (L1) English speech data compiled from the TIMIT corpus and a small portion of the Librispeech corpus [16] were used to bootstrap the training of the E2E-based model. Table 1 summarizes some basic statistics of these speech datasets.

The encoder network of the E2E-based model is composed of a 4-layer bidirectional long short-term memory (BLSTM) with 320 hidden units in each layer, while the input to the encoder network are 80-dimensional Mel-filter-bank feature vectors. In addition, the decoder network consists of a single-layer LSTM with 300 hidden units. the English canonical phone set was defined based on the CMU pronunciation dictionary.

### 3.2. Performance Evaluation

For the mispronunciation detection subtask, we follow the hierarchical evaluation structure adopted in [18], while the corresponding confusion matrix for four test conditions is illustrated in Table 2. Based on the statistics accumulated from the four test conditions, we can calculate the values of different metrics like recall (RE; TN/(FP + TN)), precision (PR; TN/(FN + TN)) and the F-1 measure (F-1; the harmonic mean of the precision and recall), so as to evaluate the performance of mispronunciation detection.

Table 1: *Statistics of the experimental speech corpora.*

| | Corpus | | subsets | Spks. | Utters. | Hrs. |
|---|---|---|---|---|---|---|
| L1 | TIMIT+ LS-sub | | Train | 989 | 27801 | 87.90 |
| | | | Dev. | 108 | 2871 | 8.83 |
| L2 | L2-ARCTIC | CP | Train | 18 | 17384 | 48.18 |
| | | | Dev. | 2 | 1962 | 4.91 |
| | | | Test | 4 | 3928 | 11.44 |
| | | MP | Train | 18 | 2697 | 7.58 |
| | | | Dev. | 2 | 300 | 0.75 |
| | | | Test | 4 | 596 | 1.75 |

Table 2: *The confusion matrix of mispronunciation detection and diagnosis task*

| Total Conditions | | Ground Truth | |
|---|---|---|---|
| | | CP | MP |
| Model Prediction | CP | True Positive (TP) | False Positive (FP) |
| | MP | False Negative (FN) | True Negative (TN) |

---

[1] https://psi.engr.tamu.edu/l2-arctic-corpus/

For the mispronunciation diagnosis subtask, we first address in those mispronounced phone labels in the text prompts of test utterances that have been correctly detected, referred to as true negative (TN; *cf.* Table 2), to calculate the diagnostic accuracy rate (DAR). Furthermore, we also analyze the performance statistics like the number (ratio) of categorical errors and non-categorical errors of the true mispronunciations (FP + TN) that we can provide correct diagnoses, respectively.

### 3.3. Experimental results

*3.3.1. Evaluations on Mispronunciation Detection*

At the outset, we assess the performance level of our proposed E2E-based method on mispronunciation detection, in relation to the cerebrated GOP-based method building on the DNN-HMM model. Specifically, the DNN component of GOP is a 5-layer time-delay neural network (TDNN) and 1,280 neurons in each layer, whose parameters were trained on the training sets of L1 and L2-CP (*cf.* Table 1). The corresponding results are shown in Table 3, where our methods were either implemented with phone-specific anti-phone modeling, viz. CTC-ATT(Anti), or with a simplified version, viz. CTC-ATT(Unk), which used a single symbol Unk instead to accommodate all non-categorical mispronunciations. Looking at Table 3, we can make at least three observations. First, our CTC-ATT(Anti) method outperforms the GOP-based method by a significant margin, demonstrating the promise of using E2E-based model structure for the mispronunciation detection subtask. Second, CTC-ATT(Anti) yields considerably better performance than CTC-ATT(Unk), which reveals that finer-grained anti-phone modeling is desirable. Third, the aforementioned methods are still far from perfect for the mispronunciation detection subtask on the L2-ARCTIC corpus.

We then set out to analyze the impacts of leveraging different model architectures on the mispronunciation detection subtask. Here apart from the hybrid CTC-Attention model (CTC-ATT), either CTC or the attention-based model (denoted by Attention for short) were investigated for this purpose. Here all the three methods were implemented with phone-specific anti-phone modeling as well (*cf.* Section 2). As can be seen from Table 4, mispronunciation detection using CTC-ATT delivers a superior F1-score than that with CTC or Attention in isolation. If we compare among CTC and Attention, it is evident that CTC stands out in performance when using recall as the evaluation metric, whereas the situation is reversed when using precision as the metric. This also confirms our anticipation that CTC-ATT is able to harness the synergistic power of CTC and ATT for use in mispronunciation detection.

*3.3.2. Evaluations on Mispronunciation Diagnosis*

In the third set of experiments, we turn to evaluating the mispronunciation diagnosis performance of our methods with different model architectures, viz. CTC-ATT, CTC and Attention. As shown in Table 5, though the DAR results of these three models still falls short of expectation, using CTC-ATT stands out in comparison to using either CTC or Attention in isolation. This also indicates that for the development of CAPT systems, the mispronunciation diagnosis subtask is even more challenging that the mispronunciation detection subtask.

Table 3: *Mispronunciation detection results of our proposed methods and the GOP-based method.*

|       | GOP   | CTC-ATT (Anti) | CTC-ATT (Unk) |
|-------|-------|----------------|---------------|
| PR(%) | 19.42 | **46.57**      | 38.99         |
| RE(%) | 52.19 | **70.28**      | 53.12         |
| F1(%) | 28.31 | **56.02**      | 44.97         |

Table 4: *Mispronunciation detection results of our proposed methods with different model structures.*

|       | CTC (Anti) | Attention (Anti) | CTC-ATT (Anti) |
|-------|------------|------------------|----------------|
| PR(%) | 41.17      | 43.89            | **46.57**      |
| RE(%) | **76.48**  | 64.54            | 70.28          |
| F1(%) | 53.52      | 52.25            | **56.02**      |

Table 5: *Mispronunciation diagnosis accuracy results (DAR%) of our proposed methods with different model structures.*

|        | CTC (Anti) | Attention (Anti) | CTC-ATT (Anti) |
|--------|------------|------------------|----------------|
| DAR(%) | 32.46      | 37.02            | **40.66**      |

Table 6: *Numbers of categorical errors and non-categorical errors of the true mispronunciations that our models can provide correct diagnoses.*

|                  | Non-categorical errors | Categorical errors |
|------------------|------------------------|--------------------|
| Ground Truth     | 100% (771)             | 100% (3,310)       |
| CTC-ATT (Unk)    | 8.4% (65)              | 19.63% (650)       |
| CTC-ATT (Anti)   | 9.4% (73)              | 33.02% (1,093)     |

As a final note, we report on a statistical analysis of the numbers (ratios) of categorical errors and non-categorical errors we could provide correct diagnoses with our two methods, viz. CTC-ATT(Anti) and CTC-ATT(Unk). Inspection of Table 6, reveals that through the use of phone-specific anti-phone modeling, viz. CTC-ATT(Anti), both the categorical and non-categorical mispronunciations can be better diagnosed than that using coarse-grained anti-phone modeling, viz. CTC-ATT(Unk).

## 4. Conclusion and Future Work

In this paper, we have presented an effective end-to-end neural modeling framework for mispronunciation detection and diagnosis (MDD), capitalizing on a hybrid CTC-Attention model structure and a novel anti-phone modeling technique. A series of empirical experiments carried on the L2-ARCTIC non-native English corpus have demonstrated its practical utility. As to future work, we are intended to investigate more sophisticated modeling techniques to characterize mispronunciations that contain non-categorial or distortion errors [19], as well as to apply and extend our methods to other L2 CAPT tasks, such as MDD for Mandarin Chinese.

# 5. References

[1] S. M. Witt and S. J. Young, "Phone-level pronunciation scoring and assessment for interactive language learning," Speech Communication, vol. 30, pp. 95–108, 2000.

[2] W. Hu, et al, "Improved mispronunciation detection with deep neural network trained acoustic models and transfer learning based logistic regression classifiers," Speech Communication, vol. 67, pp.154–166, 2015.

[3] S. Sudhakara, et al, "An improved goodness of pronunciation (GoP) measure for pronunciation evaluation with DNN-HMM system considering HMM transition probabilities," *Proceedings of the INTERSPEECH*, pp. 954–958, 2019.

[4] W. Li, et al, "Improving non-native mispronunciation detection and enriching diagnostic feedback with DNN-based speech attribute modeling," *Proceedings of the ICASSP*, pp. 6135–6139, 2016.

[5] W. Lo, S. Zhang, and H. Meng, "Automatic derivation of phonological rules for mispronunciation detection in a computer-assisted pronunciation training system," *Proceedings of the INTERSPEECH*, pp.765–768, 2010.

[6] X. Qian, F. K. Soong, and H. Meng, "Discriminative acoustic model for improving mispronunciation detection and diagnosis in computer-aided pronunciation training (CAPT)," in *International Speech Communication Association*, 2010.

[7] W. Leung, X. Liu and H. Meng, "CNN-RNN-CTC Based End-to-end Mispronunciation Detection and Diagnosis," *Proceedings of the ICASSP*, pp. 8132–8136, 2019.

[8] Y. Feng, et al, "SED-MDD: towards sentence dependent end-to-end mispronunciation detection and diagnosis," *Proceedings of the ICASSP*, pp. 3492-3496, 2020

[9] L. Chen, et al, "End-to-End neural network based automated speech scoring," *Proceedings of the ICASSP*, pp. 6234–6238, 2018.

[10] S. Mao, et al, "Unsupervised Discovery of an Extended Phoneme Set in L2 English Speech for Mispronunciation Detection and Diagnosis," *Proceedings of the ICASSP*, pp. 6244–6248, 2018.

[11] X. Li, et al, "Unsupervised discovery of non-native phonetic patterns in L2 English speech for mispronunciation detection and diagnosis" *Proceedings of the INTERSPEECH*, pp. 2554–2558, 2018.

[12] S. Watanabe, et al, "Hybrid CTC/attention architecture for end-to-end speech recognition," *IEEE Journal of Selected Topics in Signal Processing*, vol. 11, no. 8, pp. 1240–1253, 2017.

[13] O. Ronen, L. Neumeyer, and H. Franco, "Automatic detection of mispronunciation for language instruction," *Proceedings of Speech Communication and Technology*, pp. 649–652, 1997.

[14] J. K. Chorowski, et al, "Attention-based models for speech recognition," *Proceedings of NIPS*, pp. 577–585, 2015.

[15] G. Zhao, et al. "L2-ARCTIC: A Non-native English Speech Corpus," *Proceedings of the INTERSPEECH*, pp. 2783–2787, 2018.

[16] V. Panayotov, et al, "Librispeech: an asr corpus based on public domain audio books," *Proceedings of the ICASSP*, pp. 5206–5210, 2015.

[17] J. Cho, et al, "Multilingual sequence-to-sequence speech recognition: architecture, transfer learning, and language modeling," *Proceedings of the SLT*, pp. 521–527, 2018.

[18] K. Li, X. Qian and H. Meng, "Mispronunciation Detection and Diagnosis in L2 English Speech Using Multidistribution Deep Neural Networks," *IEEE/ACM Transactions on Audio, Speech, and Language Processing*, vol. 25, 193–207, 2016.

[19] S. Mao, et al, "Applying multitask learning to acoustic-phonemic model for mispronunciation detection and diagnosis in L2 English speech," *Proceedings of the ICASSP*, pp. 6254–6258, 2018.